\documentstyle[11pt,fullpage]{article}

\def\BibTeX{{\rm B\kern-.05em{\sc i\kern-.025em b}\kern-.08em
    T\kern-.1667em\lower.7ex\hbox{E}\kern-.125emX}}

\newenvironment{descit}[1]{\begin{quote} \textit{#1}}{\end{quote}}
\ifx\undefined\psfig\else \fi

\edef\psfigRestoreAt{\catcode`@=\number\catcode`@\relax}
\catcode`\@=11\relax
\newwrite\@unused
\def\typeout#1{{\let\protect\string\immediate\write\@unused{#1}}}
\def\figurepath{./}

\def\@nnil{\@nil}
\def\@empty{}
\def\@psdonoop#1\@@#2#3{}
\def\@psdo#1:=#2\do#3{\edef\@psdotmp{#2}\ifx\@psdotmp\@empty \else
    \expandafter\@psdoloop#2,\@nil,\@nil\@@#1{#3}\fi}
\def\@psdoloop#1,#2,#3\@@#4#5{\def#4{#1}\ifx #4\@nnil \else
       #5\def#4{#2}\ifx #4\@nnil \else#5\@ipsdoloop #3\@@#4{#5}\fi\fi}
\def\@ipsdoloop#1,#2\@@#3#4{\def#3{#1}\ifx #3\@nnil 
       \let\@nextwhile=\@psdonoop \else
      #4\relax\let\@nextwhile=\@ipsdoloop\fi\@nextwhile#2\@@#3{#4}}
\def\@tpsdo#1:=#2\do#3{\xdef\@psdotmp{#2}\ifx\@psdotmp\@empty \else
    \@tpsdoloop#2\@nil\@nil\@@#1{#3}\fi}
\def\@tpsdoloop#1#2\@@#3#4{\def#3{#1}\ifx #3\@nnil 
       \let\@nextwhile=\@psdonoop \else
      #4\relax\let\@nextwhile=\@tpsdoloop\fi\@nextwhile#2\@@#3{#4}}
\newread\ps@stream
\newif\ifnot@eof       
\newif\if@noisy        
\newif\if@atend        
\newif\if@psfile       
{\catcode`\%=12\global\gdef\epsf@start{
\def\epsf@PS{PS}
\def\epsf@getbb#1{%
\openin\ps@stream=#1
\ifeof\ps@stream\typeout{Error, File #1 not found}\else
   {\not@eoftrue \chardef\other=12
    \def\do##1{\catcode`##1=\other}\dospecials \catcode`\ =10
    \loop
       \if@psfile
	  \read\ps@stream to \epsf@fileline
       \else{
	  \obeyspaces
          \read\ps@stream to \epsf@tmp\global\let\epsf@fileline\epsf@tmp}
       \fi
       \ifeof\ps@stream\not@eoffalse\else
       \if@psfile\else
       \expandafter\epsf@test\epsf@fileline:. \\%
       \fi
          \expandafter\epsf@aux\epsf@fileline:. \\%
       \fi
   \ifnot@eof\repeat
   }\closein\ps@stream\fi}%
\long\def\epsf@test#1#2#3:#4\\{\def\epsf@testit{#1#2}
			\ifx\epsf@testit\epsf@start\else
\typeout{Warning! File does not start with `\epsf@start'.  It may not be a PostScript file.}
			\fi
			\@psfiletrue} 
{\catcode`\%=12\global\let\epsf@percent=
\long\def\epsf@aux#1#2:#3\\{\ifx#1\epsf@percent
   \def\epsf@testit{#2}\ifx\epsf@testit\epsf@bblit
	\@atendfalse
        \epsf@atend #3 . \\%
	\if@atend	
	   \if@verbose{
		\typeout{psfig: found `(atend)'; continuing search}
	   }\fi
        \else
        \epsf@grab #3 . . . \\%
        \not@eoffalse
        \global\no@bbfalse
        \fi
   \fi\fi}%
\def\epsf@grab #1 #2 #3 #4 #5\\{%
   \global\def\epsf@llx{#1}\ifx\epsf@llx\empty
      \epsf@grab #2 #3 #4 #5 .\\\else
   \global\def\epsf@lly{#2}%
   \global\def\epsf@urx{#3}\global\def\epsf@ury{#4}\fi}%
\def\epsf@atendlit{(atend)} 
\def\epsf@atend #1 #2 #3\\{%
   \def\epsf@tmp{#1}\ifx\epsf@tmp\empty
      \epsf@atend #2 #3 .\\\else
   \ifx\epsf@tmp\epsf@atendlit\@atendtrue\fi\fi}

\chardef\letter = 11
\chardef\other = 12

\newif \ifdebug 
\newif\ifc@mpute 
\c@mputetrue 

\let\then = \relax
\def\r@dian{pt }
\let\r@dians = \r@dian
\let\dimensionless@nit = \r@dian
\let\dimensionless@nits = \dimensionless@nit
\def\internal@nit{sp }
\let\internal@nits = \internal@nit
\newif\ifstillc@nverging
\def \Mess@ge #1{\ifdebug \then \message {#1} \fi}

{ 
	\catcode `\@ = \letter
	\gdef \nodimen {\expandafter \n@dimen \the \dimen}
	\gdef \term #1 #2 #3%
	       {\edef \t@ {\the #1}
		\edef \t@@ {\expandafter \n@dimen \the #2\r@dian}%
		\t@rm {\t@} {\t@@} {#3}%
	       }
	\gdef \t@rm #1 #2 #3%
	       {{%
		\count 0 = 0
		\dimen 0 = 1 \dimensionless@nit
		\dimen 2 = #2\relax
		\Mess@ge {Calculating term #1 of \nodimen 2}%
		\loop
		\ifnum	\count 0 < #1
		\then	\advance \count 0 by 1
			\Mess@ge {Iteration \the \count 0 \space}%
			\Multiply \dimen 0 by {\dimen 2}%
			\Mess@ge {After multiplication, term = \nodimen 0}%
			\Divide \dimen 0 by {\count 0}%
			\Mess@ge {After division, term = \nodimen 0}%
		\repeat
		\Mess@ge {Final value for term #1 of 
				\nodimen 2 \space is \nodimen 0}%
		\xdef \Term {#3 = \nodimen 0 \r@dians}%
		\aftergroup \Term
	       }}
	\catcode `\p = \other
	\catcode `\t = \other
	\gdef \n@dimen #1pt{#1} 
}

\def \Divide #1by #2{\divide #1 by #2} 

\def \Multiply #1by #2
       {{
	\count 0 = #1\relax
	\count 2 = #2\relax
	\count 4 = 65536
	\Mess@ge {Before scaling, count 0 = \the \count 0 \space and
			count 2 = \the \count 2}%
	\ifnum	\count 0 > 32767 
	\then	\divide \count 0 by 4
		\divide \count 4 by 4
	\else	\ifnum	\count 0 < -32767
		\then	\divide \count 0 by 4
			\divide \count 4 by 4
		\else
		\fi
	\fi
	\ifnum	\count 2 > 32767 
	\then	\divide \count 2 by 4
		\divide \count 4 by 4
	\else	\ifnum	\count 2 < -32767
		\then	\divide \count 2 by 4
			\divide \count 4 by 4
		\else
		\fi
	\fi
	\multiply \count 0 by \count 2
	\divide \count 0 by \count 4
	\xdef \product {#1 = \the \count 0 \internal@nits}%
	\aftergroup \product
       }}

\def\r@duce{\ifdim\dimen0 > 90\r@dian \then   
		\multiply\dimen0 by -1
		\advance\dimen0 by 180\r@dian
		\r@duce
	    \else \ifdim\dimen0 < -90\r@dian \then  
		\advance\dimen0 by 360\r@dian
		\r@duce
		\fi
	    \fi}

\def\Sine#1%
       {{%
	\dimen 0 = #1 \r@dian
	\r@duce
	\ifdim\dimen0 = -90\r@dian \then
	   \dimen4 = -1\r@dian
	   \c@mputefalse
	\fi
	\ifdim\dimen0 = 90\r@dian \then
	   \dimen4 = 1\r@dian
	   \c@mputefalse
	\fi
	\ifdim\dimen0 = 0\r@dian \then
	   \dimen4 = 0\r@dian
	   \c@mputefalse
	\fi
	\ifc@mpute \then
		\divide\dimen0 by 180
		\dimen0=3.141592654\dimen0
		\dimen 2 = 3.1415926535897963\r@dian 
		\divide\dimen 2 by 2 
		\Mess@ge {Sin: calculating Sin of \nodimen 0}%
		\count 0 = 1 
		\dimen 2 = 1 \r@dian 
		\dimen 4 = 0 \r@dian 
		\loop
			\ifnum	\dimen 2 = 0 
			\then	\stillc@nvergingfalse 
			\else	\stillc@nvergingtrue
			\fi
			\ifstillc@nverging 
			\then	\term {\count 0} {\dimen 0} {\dimen 2}%
				\advance \count 0 by 2
				\count 2 = \count 0
				\divide \count 2 by 2
				\ifodd	\count 2 
				\then	\advance \dimen 4 by \dimen 2
				\else	\advance \dimen 4 by -\dimen 2
				\fi
		\repeat
	\fi		
			\xdef \sine {\nodimen 4}%
       }}

\def\Cosine#1{\ifx\sine\UnDefined\edef\Savesine{\relax}\else
		             \edef\Savesine{\sine}\fi
	{\dimen0=#1\r@dian\multiply\dimen0 by -1
	 \advance\dimen0 by 90\r@dian
	 \Sine{\nodimen 0}
	 \xdef\cosine{\sine}
	 \xdef\sine{\Savesine}}}	      

\def\psdraft{
	\def\@psdraft{0}
}
\def\psfull{
	\def\@psdraft{100}
}

\psfull

\newif\if@draftbox
\def\psnodraftbox{
	\@draftboxfalse
}
\@draftboxtrue

\newif\if@prologfile
\newif\if@postlogfile
\def\pssilent{
	\@noisyfalse
}
\def\psnoisy{
	\@noisytrue
}
\psnoisy
\newif\if@bbllx
\newif\if@bblly
\newif\if@bburx
\newif\if@bbury
\newif\if@height
\newif\if@width
\newif\if@rheight
\newif\if@rwidth
\newif\if@angle
\newif\if@clip
\newif\if@verbose
\newif\if@scale
\def\@p@@sclip#1{\@cliptrue}

\def\@p@@sfile#1{\def\@p@sfile{null}%
	        \openin1=#1
		\ifeof1\closein1%
		       \openin1=\figurepath#1
			\ifeof1\typeout{Error, File #1 not found}
			   \if@bbllx\if@bblly\if@bburx\if@bbury
			      \def\@p@sfile{#1}%
			   \fi\fi\fi\fi
			\else\closein1
			    \edef\@p@sfile{\figurepath#1}%
                        \fi%
		 \else\closein1%
		       \def\@p@sfile{#1}%
		 \fi}
\def\@p@@sfigure#1{\def\@p@sfile{null}%
	        \openin1=#1
		\ifeof1\closein1%
		       \openin1=\figurepath#1
			\ifeof1\typeout{Error, File #1 not found}
			   \if@bbllx\if@bblly\if@bburx\if@bbury
			      \def\@p@sfile{#1}%
			   \fi\fi\fi\fi
			\else\closein1
			    \def\@p@sfile{\figurepath#1}%
                        \fi%
		 \else\closein1%
		       \def\@p@sfile{#1}%
		 \fi}

\def\@p@@sbbllx#1{
		\@bbllxtrue
		\dimen100=#1
		\edef\@p@sbbllx{\number\dimen100}
}
\def\@p@@sbblly#1{
		\@bbllytrue
		\dimen100=#1
		\edef\@p@sbblly{\number\dimen100}
}
\def\@p@@sbburx#1{
		\@bburxtrue
		\dimen100=#1
		\edef\@p@sbburx{\number\dimen100}
}
\def\@p@@sbbury#1{
		\@bburytrue
		\dimen100=#1
		\edef\@p@sbbury{\number\dimen100}
}
\def\@p@@sheight#1{
		\@heighttrue
		\dimen100=#1
   		\edef\@p@sheight{\number\dimen100}
}
\def\@p@@swidth#1{
		\@widthtrue
		\dimen100=#1
		\edef\@p@swidth{\number\dimen100}
}
\def\@p@@srheight#1{
		\@rheighttrue
		\dimen100=#1
		\edef\@p@srheight{\number\dimen100}
}
\def\@p@@srwidth#1{
		\@rwidthtrue
		\dimen100=#1
		\edef\@p@srwidth{\number\dimen100}
}
\def\@p@@sangle#1{
		\@angletrue
		\edef\@p@sangle{#1} 
}
\def\@p@@ssilent#1{ 
		\@verbosefalse
}
\def\@p@@sscale#1{
		\def\@p@scale{#1}
		\@scaletrue
}
\def\@p@@sprolog#1{\@prologfiletrue\def\@prologfileval{#1}}
\def\@p@@spostlog#1{\@postlogfiletrue\def\@postlogfileval{#1}}
\def\@cs@name#1{\csname #1\endcsname}
\def\@setparms#1=#2,{\@cs@name{@p@@s#1}{#2}}
\def\ps@init@parms{
		\@bbllxfalse \@bbllyfalse
		\@bburxfalse \@bburyfalse
		\@heightfalse \@widthfalse
		\@rheightfalse \@rwidthfalse
		\@scalefalse
		\def\@p@sbbllx{}\def\@p@sbblly{}
		\def\@p@sbburx{}\def\@p@sbbury{}
		\def\@p@sheight{}\def\@p@swidth{}
		\def\@p@srheight{}\def\@p@srwidth{}
		\def\@p@sangle{0}
		\def\@p@sfile{}
		\def\@p@scost{10}
		\def\@sc{}
		\@prologfilefalse
		\@postlogfilefalse
		\@clipfalse
		\if@noisy
			\@verbosetrue
		\else
			\@verbosefalse
		\fi
}
\def\parse@ps@parms#1{
	 	\@psdo\@psfiga:=#1\do
		   {\expandafter\@setparms\@psfiga,}}
\newif\ifno@bb
\def\bb@missing{
	\if@verbose{
		\typeout{psfig: searching \@p@sfile \space  for bounding box}
	}\fi
	\no@bbtrue
	\epsf@getbb{\@p@sfile}
        \ifno@bb \else \bb@cull\epsf@llx\epsf@lly\epsf@urx\epsf@ury\fi
}	
\def\bb@cull#1#2#3#4{
	\dimen100=#1 bp\edef\@p@sbbllx{\number\dimen100}
	\dimen100=#2 bp\edef\@p@sbblly{\number\dimen100}
	\dimen100=#3 bp\edef\@p@sbburx{\number\dimen100}
	\dimen100=#4 bp\edef\@p@sbbury{\number\dimen100}
	\no@bbfalse
}

\newdimen\p@intvaluex
\newdimen\p@intvaluey
\newdimen\@ffsetvalue
\newdimen\x@ffsetvalue
\newdimen\y@ffsetvalue

\def\compute@offset#1#2{{\dimen0=#1 sp\dimen1=#2 sp
			\advance\dimen1 by -\dimen0
			\dimen1=\sine\dimen1
			\dimen0=\cosine\dimen1
			\ifdim\dimen0<0sp \dimen1=0sp \fi
			\global\@ffsetvalue=\dimen1}}

\def\rotate@#1#2{{\dimen0=#1 sp\dimen1=#2 sp
		  \global\p@intvaluex=\cosine\dimen0
		  \dimen3=\sine\dimen1
		  \global\advance\p@intvaluex by -\dimen3
		  \global\p@intvaluey=\sine\dimen0
		  \dimen3=\cosine\dimen1
		  \global\advance\p@intvaluey by \dimen3
		  }}
\def\compute@bb{
		\no@bbfalse
		\if@bbllx \else \no@bbtrue \fi
		\if@bblly \else \no@bbtrue \fi
		\if@bburx \else \no@bbtrue \fi
		\if@bbury \else \no@bbtrue \fi
		\ifno@bb \bb@missing \fi
		\ifno@bb \typeout{FATAL ERROR: no bb supplied or found}
			\no-bb-error
		\fi
		\if@angle 
			\Sine{\@p@sangle}\Cosine{\@p@sangle}
			\compute@offset{\@p@sbblly}{\@p@sbbury}
			\x@ffsetvalue=\@ffsetvalue
			\compute@offset{\@p@sbburx}{\@p@sbbllx}
			\y@ffsetvalue=\@ffsetvalue

			\rotate@{\@p@sbbllx}{\@p@sbblly}
			\advance\p@intvaluex by -\x@ffsetvalue
			\advance\p@intvaluey by -\y@ffsetvalue
			\edef\@p@sbbllx{\number\p@intvaluex}
			\edef\@p@sbblly{\number\p@intvaluey}

			\rotate@{\@p@sbburx}{\@p@sbbury}
			\advance\p@intvaluex by \x@ffsetvalue
			\advance\p@intvaluey by \y@ffsetvalue
			\edef\@p@sbburx{\number\p@intvaluex}
			\edef\@p@sbbury{\number\p@intvaluey}
			{
			 \count0=\@p@sbbllx \count1=\@p@sbblly
		 	 \count2=\@p@sbburx \count3=\@p@sbbury
			 \dimen0=\@p@sbbllx sp\dimen1=\@p@sbblly sp
		 	 \dimen2=\@p@sbburx sp\dimen3=\@p@sbbury sp
			 \dimen203=\dimen2 \advance\dimen203 by -\dimen0
			 \dimen204=\dimen3 \advance\dimen204 by -\dimen1
			 \ifdim\dimen203<0sp 
			      \count203=\count2 \count2=\count0 
			      \count0=\count203 
			      \global\edef\@p@sbbllx{\number\count0}
			      \global\edef\@p@sbburx{\number\count2}
			 \fi
			 \ifdim\dimen204<0sp 
			       \count204=\count3
			       \count3=\count1
			       \count1=\count204
			       \global\edef\@p@sbblly{\number\count1}
			       \global\edef\@p@sbbury{\number\count3}
			 \fi
			}
		\fi
		\count203=\@p@sbburx
		\count204=\@p@sbbury
		\advance\count203 by -\@p@sbbllx
		\advance\count204 by -\@p@sbblly
		\edef\@bbw{\number\count203}
		\edef\@bbh{\number\count204}
}
\def\in@hundreds#1#2#3{\count240=#2 \count241=#3
		     \count100=\count240	
		     \divide\count100 by \count241
		     \count101=\count100
		     \multiply\count101 by \count241
		     \advance\count240 by -\count101
		     \multiply\count240 by 10
		     \count101=\count240	
		     \divide\count101 by \count241
		     \count102=\count101
		     \multiply\count102 by \count241
		     \advance\count240 by -\count102
		     \multiply\count240 by 10
		     \count102=\count240	
		     \divide\count102 by \count241
		     \count200=#1\count205=0
		     \count201=\count200
			\multiply\count201 by \count100
		 	\advance\count205 by \count201
		     \count201=\count200
			\divide\count201 by 10
			\multiply\count201 by \count101
			\advance\count205 by \count201
		     \count201=\count200
			\divide\count201 by 100
			\multiply\count201 by \count102
			\advance\count205 by \count201
		     \edef\@result{\number\count205}
}
\def\@ScaleInHundreds#1{
		\in@hundreds{#1}{\@p@scale}{100}
		\edef#1{\@result}
}
\def\compute@wfromh{
		\in@hundreds{\@p@sheight}{\@bbw}{\@bbh}
		\edef\@p@swidth{\@result}
}
\def\compute@hfromw{
		\in@hundreds{\@p@swidth}{\@bbh}{\@bbw}
		\edef\@p@sheight{\@result}
}
\def\compute@handw{
		\if@height 
			\if@width
			\else
				\compute@wfromh
			\fi
		\else 
			\if@width
				\compute@hfromw
			\else
				\edef\@p@sheight{\@bbh}
				\edef\@p@swidth{\@bbw}
			\fi
		\fi
}
\def\compute@resv{
		\if@rheight \else \edef\@p@srheight{\@p@sheight} \fi
		\if@rwidth \else \edef\@p@srwidth{\@p@swidth} \fi
}
\def\compute@sizes{
	\compute@bb
	\compute@handw
	\compute@resv
}
\def\psfig#1{\vbox {
	\ps@init@parms
	\parse@ps@parms{#1}
	\compute@sizes
	\if@scale
                \if@verbose
                        \typeout{psfig: scaling by \@p@scale}
                \fi
                \@ScaleInHundreds{\@p@swidth}
                \@ScaleInHundreds{\@p@sheight}
                \@ScaleInHundreds{\@p@srwidth}
                \@ScaleInHundreds{\@p@srheight}
        \fi
	\ifnum\@p@scost<\@psdraft{
		\if@verbose{
			\typeout{psfig: including \@p@sfile \space }
		}\fi
		\special{ps::[begin] 	\@p@swidth \space \@p@sheight \space
				\@p@sbbllx \space \@p@sbblly \space
				\@p@sbburx \space \@p@sbbury \space
				startTexFig \space }
		\if@angle
			\special {ps:: \@p@sangle \space rotate \space} 
		\fi
		\if@clip{
			\if@verbose{
				\typeout{(clip)}
			}\fi
			\special{ps:: doclip \space }
		}\fi
		\if@prologfile
		    \special{ps: plotfile \@prologfileval \space } \fi
		\special{ps: plotfile \@p@sfile \space }
		\if@postlogfile
		    \special{ps: plotfile \@postlogfileval \space } \fi
		\special{ps::[end] endTexFig \space }
		\vbox to \@p@srheight true sp{
			\hbox to \@p@srwidth true sp{
				\hss
			}
		\vss
		}
	}\else{
		\if@draftbox{		
			\hbox{\fbox{\vbox to \@p@srheight true sp{
			\vss
			\hbox to \@p@srwidth true sp{ \hss \@p@sfile \hss }
			\vss
			}}}
		}\else{
			\vbox to \@p@srheight true sp{
			\vss
			\hbox to \@p@srwidth true sp{\hss}
			\vss
			}
		}\fi

	}\fi
}}
\def\psglobal{\typeout{psfig: PSGLOBAL is OBSOLETE; use psprint -m instead}}
\psfigRestoreAt

\begin{document}
\pagestyle{myheadings}
\markboth{Not for distribution or attribution; For review purposes only.}{Not for distribution or attribution; For review purposes only.}
\markright{Not for distribution or attribution; For review purposes only.}

\title{PIPE: Personalizing Recommendations\\
via Partial Evaluation}

\author{Naren Ramakrishnan\\
Department of Computer Science\\
Virginia Tech, Blacksburg, VA 24061\\
Email: naren@cs.vt.edu}

\date{}
\maketitle


\hrule

\begin{abstract}

\noindent
It is shown that personalization of web content can be advantageously viewed
as a form of partial evaluation --- a technique well known in the programming
languages community. The basic idea is to model a
recommendation space as a program, then partially evaluate this
program with respect to user preferences (and features) to obtain 
specialized content.
This technique supports both content-based and collaborative approaches,
and is applicable to a range of applications that require automatic
information integration from multiple web sources. The
effectiveness of this methodology is illustrated by two example applications ---
(i) personalizing content for visitors to the Blacksburg Electronic
Village ({\tt http://www.\hskip0ex bev.\hskip0ex net}), and (ii)
locating and selecting scientific software on the Internet.
The scalability of this technique is demonstrated by its ability
to interface with online web ontologies that index thousands
of web pages.
\end{abstract}

\newpage

\section{Introduction}
Personalization of web content constitutes one of the fastest
growing segments of the Internet economy today \cite{specissue,shapiro}.
It helps to retain customers, reduces information overload, and
enables mass customization in E-commerce \cite{pine}. 

Two main approaches have been proposed to conduct personalization.
The simplest are web search engines and the information filtering schemes
which use {\it content-based techniques} to
alleviate information overload. They, however, harness only a small fraction of the 
indexable web (one study
estimates this to be $< 30\%$ \cite{lawrence}), and still require users to sift through
a multitude of results to determine relevant selections. The low coverage
of search engines is attributed to at least two reasons: (i) a majority of
web pages are dynamically generated \cite{db-www} (and hence not directly
accessible via hyperlinks), and (ii) lack of sophisticated conceptual
models for web information retrieval. At the other end of the spectrum, {\it collaborative
filtering techniques} mine user access patterns, web logs, preferences, and profiles to precisely
tailor the content provided (``Since you liked `Sense and Sensibility,' you might
be interested in `Pride and Prejudice' too'') at specific sites. As businesses race to provide
comprehensive experiences to web visitors, various combinations of these two approaches
are used. This has spawned a multimillion dollar industry (NetPerceptions, Imana etc.)
that provides custom-built personalization
solutions for individual client specifications.
See the web portal {\tt www.\hskip0ex personalization.\hskip0ex com} for information
on all aspects of this industry.

In this article, a  customizable methodology called PIPE is presented that can be
used to design personalization systems for a specific application (involving a
collection of sites). PIPE allows the incorporation of both content-based and collaborative filtering
techniques.  It supports information integration,
varying levels of input by web visitors, and facilitates ease of construction by a skilled
systems engineer. For example, a designer wishing to construct a personalization facility for
`wines' can model various web resources (pertaining to this domain) using PIPE and create
a facility that customizes content for visitors, based on wine preferences and attributes.

The rest of the article is organized as follows. Section 2 introduces
{\it partial evaluation} --- the key concept behind the methodology of PIPE.
Sections 3 and 4 develop this idea further and present various schemes that extend it
to large domains. Section 5 presents two case studies implemented using this
framework. The evaluation of both these implementations are undertaken next, in Section 6. 
Section 7 discusses various aspects of the PIPE framework, its evaluation, and how it
relates to other approaches to personalization.

\section{Personalization is Partial Evaluation}

The central contribution of this article is to model personalization by the programmatic
notion of partial evaluation. Partial evaluation is a technique to specialize
programs, given incomplete information about their input \cite{jones}.
The methodology presented here models a web site
as a program (which abstracts the underlying schema of organization), partially evaluates 
the program
with respect to user input, and recreates a personalized
web site from the specialized program.

The input to a partial evaluator
is a program and (some) static information about its arguments. Its
output is a specialized version of this program (typically in the same
language),
that uses the static information to `pre-compile' as many operations
as possible. A simple example is how the C function {\tt pow}
can be specialized to create a new function, say
{\tt pow2}, that computes the square of an integer. Consider for example,
the definition of a {\tt pow}er function shown in the left part of Fig.~\ref{pe}
(grossly simplified for presentation purposes).
If we knew that a particular user will utilize it 
only for computing squares of
integers, we could specialize it (for that user) to produce the {\tt pow2} function.
Thus, {\tt pow2} is obtained automatically (not by a human programmer)
from {\tt pow} by precomputing all expressions that involve {\tt exponent},
unfolding the for-loop and by various other compiler transformations such as
copy propagation and forward substitution.
Its benefit is obvious when we consider a higher-level loop
that would invoke {\tt pow} repeatedly for computing, say, the squares of
all integers from 1 through 100. Partial evaluation is now used in a wide
variety of applications (scientific computing, database systems etc.)
to achieve speedup in highly parameterized environments. Automatic
program specializers are available for C, FORTRAN, PROLOG, LISP, and several other important
languages. The interested reader is referred to \cite{jones} for a good introduction.
While the traditional motivation for using partial evaluation is to achieve speedup
and/or remove interpretation overhead, it can also be viewed as a technique
to simplify program presentation, by removing inapplicable, unnecessary,
and `uninteresting' information (based on user criteria) from a program.\\

\begin{figure}
\centering
\begin{tabular}{|l|l|} \hline
{\tt int pow(int base, int exponent) \{} & {\tt int pow2(int base) \{} \\
\,\,\,\,\,{\tt int prod = 1;} & \,\,\,\,\,{\tt return (base * base)} \\
\,\,\,\,\,{\tt for (int i=0;i<exponent;i++)} &  \} \\
\,\,\,\,\,\,\,\,\,\,{\tt prod = prod * base;} & \\
\,\,\,\,\,{\tt return (prod);} & \\
\} & \\
\hline
\end{tabular}
\caption{Illustration of the partial evaluation technique.
A general purpose {\tt pow}er function written in C (left) and
its specialized version (with {\tt exponent} = 2) to handle squares
(right).
Such specializations are performed automatically by partial evaluators
such as C-Mix \cite{jones}.}
\label{pe}
\end{figure}

\begin{figure}
\centering
\begin{tabular}{cc}
& \mbox{\psfig{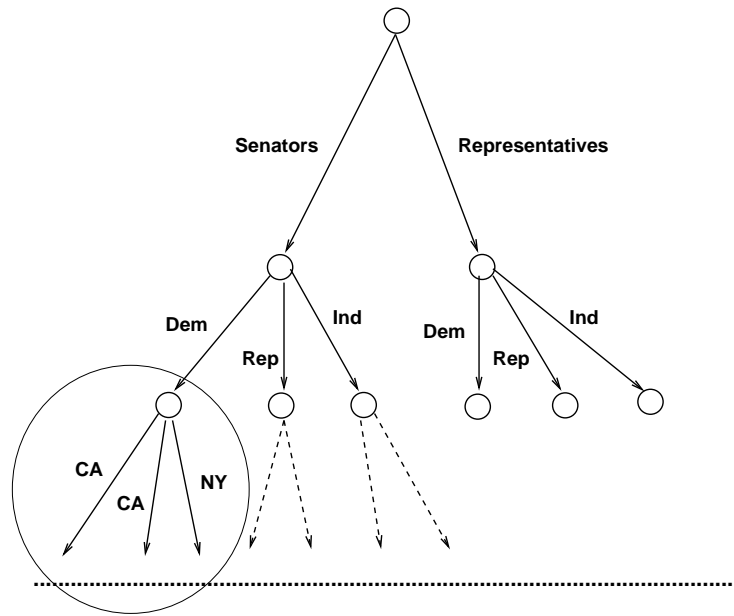}}\\
\end{tabular}
\caption{Hypothetical web site organization modeling information about US Senators and 
Representatives. Only the first few levels of the site are shown; the lower levels
can be visualized as modeling individual precincts of politicians, the bills 
they sponsor, constituencies, their addresses, interests, education etc. The labels
on edges represent choices and selections made by a navigator. The circled region
indicates the result of personalization for the input `Democratic Senators.'}
\label{senator}
\end{figure}

\noindent
{\it Example 1:} We now present a simple example to
illustrate how a web site can be abstracted as a program for partial evaluation. 
Consider a congressional web site,
organized in a hierarchical fashion, that provides information about US Senators, Representatives,
their party, precinct, and state affiliations (Fig.~\ref{senator}). Later examples
will remove this restriction to hierarchical sites.
The individual web pages in Fig.~\ref{senator} are denoted
by the nodes (circles), and the links are assumed to be tagged via some labeling mechanism.
Such labels can be obtained from the text anchoring the hyperlinks via
`{\tt <a href>}s' in the web pages, or from XML tags \cite{xmltour}. A web crawler
employing a depth-first search can then be used to obtain a program,
that models the links in a way that the interpretation of the program refers to the organization of
information in the web sources. For example, the data in Fig.~\ref{senator} produces the
program (the line numbers are shown for ease of reference):

\begin{center}
\begin {tabular}{|p{4.2in}|}\hline
\begin{verbatim}
1:  if (Senators)
2:     if (Dem)
3:        if (CA)
4:          .....
5:        else if (NY)
6:          .....
7:     else if (Rep)
8:        ....
9:     ...
10: else if (Representatives)
11:    if (Dem)
12:       ....
\end{verbatim}
\\\hline
\end {tabular}
\end{center}
\vspace{.075in}

\noindent
where the link labels are represented as program variables.
The mutually-exclusive dichotomies of links at individual nodes (e.g. `A Democrat
cannot be a Republican') are modeled by {\tt else if}s\footnote{This issue is addressed in
more detail in Section 7.}.
Notice that  while the program only models the organization of the web site, other 
textual information at each of the internal nodes can be stored/indexed alongside by associating
augmented data structures with the program variables.
Furthermore, at the `leaves' (i.e., the innermost sections of the program), variable
assignments corresponding to the individual URLs of the Senator/Representative home pages can
be stored. 

Assume that a user is interested in personalizing the web site to provide information only
about `Democratic Senators.' This is easily achieved by partially evaluating the above program
with respect to the variables {\tt Senators} and {\tt Dem} (setting them to $1$). This produces
the simplified program:

\begin{center}
\begin {tabular}{|p{4.2in}|}\hline
\begin{verbatim}
3:        if (CA)
4:          .....
5:        else if (NY)
6:          .....
\end{verbatim}
\\\hline
\end {tabular}
\end{center}
\vspace{.075in}

\noindent
which can be used to recreate web pages, thus yielding personalized web content (shown by
the circular region in Fig.~\ref{senator}).
The flexibility of this approach is that it allows personalization even
when variable values for certain level(s) are available, but not for level(s)
higher in the hierarchy.
For example, if the user desires information
about a NY
politician (but is unsure whether he/she is a Senator or Representative or a 
Democrat/Republican/Independent), then a partially evaluated output (with respect
to {\tt NY} and setting other variables such as {\tt CA} to zero) will simplify the 
lower levels of the hierarchy, yielding:

\begin{center}
\begin {tabular}{|p{4.2in}|}\hline
\begin{verbatim}
1:  if (Senators)
2:     if (Dem)
6:          .....
7:     else if (Rep)
8:        ....
9:     ...
10: else if (Representatives)
11:    if (Dem)
12:       ....
\end{verbatim}
\\\hline
\end {tabular}
\end{center}
\vspace{.075in}

\noindent
The approach is thus responsive to varying levels of user input, ranging from no information
(wherein the original program will be reproduced in its entirety)
to choices that completely determine the end web page(s). \hrulefill \\

\section{Mining Semi-Structured Data}
The simplistic approach presented in {\it Example 1}
will be infeasible for realistic web sites, which are
not strictly hierarchical, and best abstracted by
`semi-structured' data models, a term that has come to denote implicit,
loose, irregular, and
constantly evolving schema of information. 
To scale this methodology for semi-structured data, we propose the application of data mining
techniques that extract compressed schema from web sites.

\begin{figure}
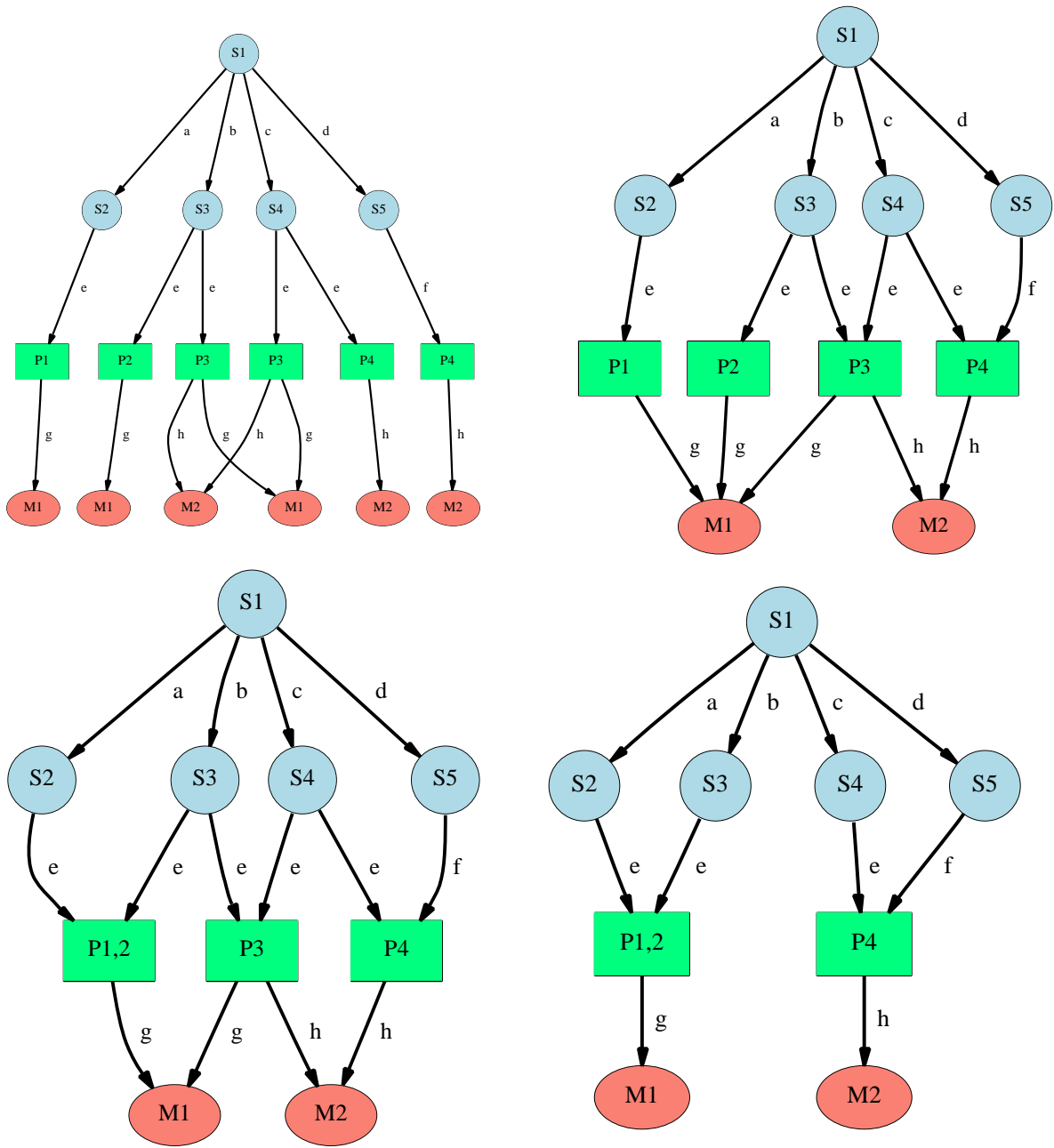

\begin{tabular}{cc}
& \mbox{\psfig{figure=ex12.epsi,width=2.8in}}
\end{tabular}
\begin{tabular}{cc}
& \mbox{\psfig{figure=ex2.epsi,width=2.8in}}
\end{tabular}
\begin{tabular}{cc}
& \mbox{\psfig{figure=ex3.epsi,width=2.8in}}
\end{tabular}
\begin{tabular}{cc}
& \mbox{\psfig{figure=ex4.epsi,width=2.8in}}
\end{tabular}
\caption{Four stages in mining schema from a semi-structured data source.
The input is assumed to be a graph with labeled and directed edges (top left).
Commonalities encountered in tree-building are factored first (top right). At this stage,
multiple internal nodes may possess the same input and output labels (for example, {\tt P1}
and {\tt P2}). The algorithm then proceeds to type
the data, thus collapsing {\tt P1} and {\tt P2} (bottom left). Finally, nodes are allowed
to belong to multiple types, rendering {\tt P3} to be redundant (bottom right).}
\label{4nest}
\end{figure}

Various graph-based models have been proposed to model semi-structured data which, again, use
directed labeled arcs to model the connection between web pages and between web sites.
The data mining techniques that operate on such models obey the `approximation model' 
of data mining. That is, they start
with an accurate (exact) model of the data and deliberately introduce approximations, in
the hope of finding some latent/hidden structure to the data. 
We illustrate the basic idea using the approximation model of Nestorov et al. \cite{nestorov},
which treats web pages as atomic objects, and models the links between web pages as
relations between the atomic objects.\\

\noindent
{\it Example 2:} Consider the hypothetical web site depicted in the top left part of
Fig.~\ref{4nest}. The individual web pages
are denoted by {\tt M1}, {\tt M2} etc., the pre-leaf nodes by {\tt P1}, {\tt P2}
etc., while the other internal nodes are represented as {\tt S1}, {\tt S2} etc. Notice again
that the links
are assumed to be tagged via some labeling mechanism.
The first step in extracting structure from the
web site is to proceed to {\it type} the data i.e., determine the
minimum number of entities needed to model the web schema.
For example, the {\tt S2} node in the top left part of Fig.~\ref{4nest} can
be typed as: \\

\noindent
{\tt S2(Y) :- S1(X), link(X,Y,'a'), P1(Z), link(Y,Z,'e')}\\

\noindent
which indicates that
it is reachable from {\tt S1} (using the {\tt a} tag), and has a link to {\tt P1} (via the
label {\tt e}). Such a typing (expressed in the form of a logic program)
might not yield any compression to the original data, so various
approximations and simplifications are employed to reduce its size before partial evaluation.
We first identify commonalities due to encountering the same page multiple times (top right
of Fig.~\ref{4nest}). This is easily achieved by using a hash indexed by page URL in
the web crawler.
Next, the algorithm of Nestorov et al. \cite{nestorov} uses
program-theoretic techniques to find the minimal set of types necessary to accurately represent
the original data. For example, {\tt P1} and {\tt P2} have the same input
and output labels (to the same page), and can be compressed into a single
type {\tt P1,2}, by computing the greatest fixed-point of the logic program (see \cite{nestorov}).
And finally, allowing one type to be expressed as the superposition of
multiple other types helps further reduce the size of the logic program.  In this
case, {\tt P3} can be subsumed by a combination of {\tt P1,2} and {\tt P4}.
The end-result of 
this process
(see bottom right of Fig.~\ref{4nest}) is a succinct
schema that can be used for personalization.  
The cost of the mining algorithm is double-quadratic
in the size of the web site (pre-leaf nodes). For web sites that are purely hierarchical
and that do not contain cycles, more simplifications are available that enable efficient
implementations of the mining algorithm \cite{nestorov}. \hrulefill\\

\section{Information Integration}
The methodology described thus far is content-based, works at the level of web site
organization, and does not mine/model the textual content or formatting
within individual pages, beyond associating
them with the appropriate nodes in the graph. For example, if a web page has two links {\tt L1}
and {\tt L2} to other pages, the text anchoring {\tt L1} (upto the start of {\tt L2}) is 
associated with
the node corresponding to {\tt L1} in the graph, and so on. While data mining techniques
are available that deal with textual information, we do not employ them in our study and
implementations. We thus restrict our studies to web sites where most of
the information content to be personalized is found at the `leaves' of the structure tree.
In addition, the ideas presented above are restricted to a single site. It is well understood
that to provide compelling personalization scenarios, information needs to be integrated
from multiple web sites and other sources of information, such as recommender systems \cite{ira,grouplens,phoaks} and
topic-specific cross-indices. Recommender systems, as introduced in Section 1, make selections of artifacts
by mining profiles of customer choices and buying patterns. 
Topic-specific indices
provide ontologies and taxonomies by cross-referencing information from multiple sites (e.g.
the Yahoo taxonomy). \\

\noindent
{\it Example 3:} Consider personalizing stock quotes for potential investors. The Yahoo!
Finance Cross-Index at [{\tt quote.yahoo.com}] provides a ticker symbol lookup for 
stock charts, financial statistics, and links to company profiles.
It is easy to model and personalize
this site by the techniques presented in previous sections. The program obtained 
could be
partially evaluated with respect to ticker symbol to yield specialized information. 
However, what if
the user does not know the ticker symbol, but has access to only the company name? What
if he/she desires to index based on recommendations from an online brokerage? The key
issue thus is to provide support for information integration from multiple web resources.
This entails the inherent inconsistencies and uncertainties
in representing and modeling textual labels.
The online brokerage might refer to its recommendations by company name (e.g. `Microsoft'),
while the Yahoo! cross-index uses the ticker symbol (`MSFT'). More seriously, financial
terms carry with them the twin idiosyncrasies of synonymy and polysemy. For example, `Investments'
referred to in one web site might be listed as `Ventures' in another, which might mean something
completely different in a non-financial setting. 

We now outline possible solutions to these issues. The choices made by an individual
recommender system can be modeled as statements in a program that abstract the control
flow of the selection algorithm. For example, in the financial
setting above, a special function can be written that takes as input the current 
user profile and returns a ticker symbol recommendation. This function can
be called from a {\tt main()} routine, which can then use the resulting ticker symbol
to set variables for the program obtained by mining the Finance Cross-Index. In addition,
the issue of synonymy can be addressed by introducing additional assertions such as:

\begin{center}
\begin {tabular}{|p{4.2in}|}\hline
\begin{verbatim}
   if (MSFT)
      Microsoft= TRUE;
   if (Microsoft)
      MSFT = TRUE;
\end{verbatim}
\\\hline
\end {tabular}
\end{center}
\vspace{.075in}

\noindent
at the beginning of the composite program, thus abstracting the task models underlying the
application.
This is the most domain-specific part of
the methodology and cannot be easily automated (this aspect is discussed in more detail
later in the paper). The literature on information integration proposes various
solutions to this problem, notably wrappers and mediator-based schemes \cite{db-www}. \hrulefill 

\section{Case Studies}
The above three aspects of (i) partial evaluation, (ii) mining
semi-structured data, and (iii) information integration  form the basis of the PIPE
methodology for personalization. 
PIPE serves as a customizable framework that can
compose individual content-based and collaborative engines
to form full-fledged personalization systems in a specific domain.
To the best of the author's knowledge, there exists no comparable methodology
for designing personalization systems, though similar
architectures are available for other aspects of information capture and access
\cite{rus}. We now summarize the various steps of this methodology. 

The first step is to identify the different `starting points' for personalization --- a
domain specific consideration. The schema in these various sites should be modeled by
labeled graphs involving semi-structured data. The second step is to extract typing rules
from each of the site structures by the mining algorithm. The third step is to merge
the diverse schema into a composite program, taking care to ensure that 
entities referred to in different ways by individual
web sources are correctly merged together. We refer to the information
space represented by the composite program as a `recommendation space.'
These steps constitute the {\it off-line}
aspect of the methodology and need be performed only once for a specific implementation.
The final step is the online aspect of partially evaluating the composite program and
reconstructing the original information from the specialized program. 
We present empirical evidence for the effectiveness of this approach by application
to two diverse domains:
\begin{itemize}
\item Creating personalized web pages for scientists and engineers
who are trying to locate software on the Internet, and
\item Delivering web-based tourist information for visitors to
the Blacksburg Electronic Village ({\tt http://www.bev.net}) (BEV).
\end{itemize}
The common denominator among these applications is their strong emphasis on
multiple information resources (needed to achieve the desired effect),
the heterogeneity of the sources, and the desire to provide integrated
support for interesting personalization scenarios.

\subsection{Personalizing Content for Scientists and Engineers}
\begin{figure}
\centering
\begin{tabular}{cc}
& \mbox{\psfig{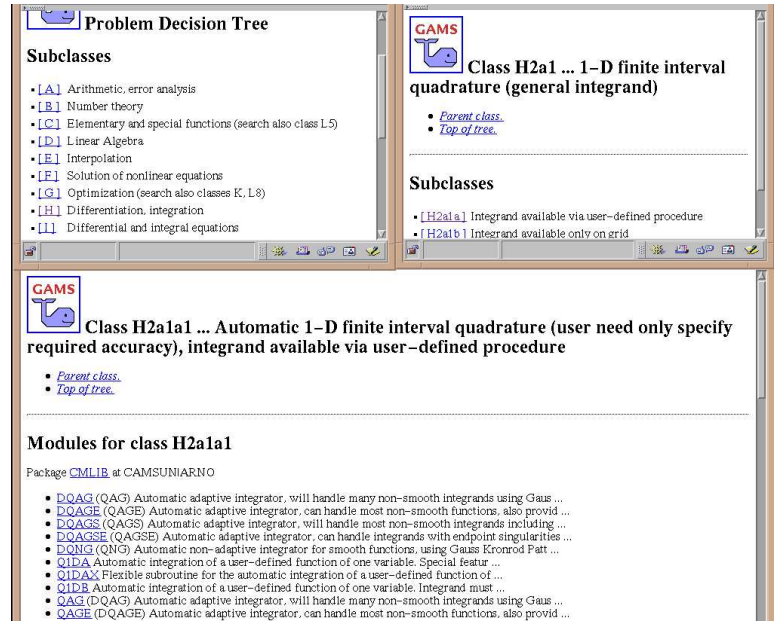}}\\
\end{tabular}
\caption{Snapshot of the GAMS Interface ({\tt http://\hskip0ex gams.\hskip0ex nist.\hskip0ex gov})
at three levels of the hierarchy.}
\label{gams1}
\end{figure}

The PIPE methodology has been used in the context of creating
personalized recommendations about mathematical and scientific software
on the web. One of the main research issues here is understanding the
fundamental
processes by which knowledge about scientific problems is created,
validated and communicated. Designing a personalization system for
this domain involves at least three different sources:

\begin{itemize}

\item {\bf Web-based software repositories:} In our study, we chose
Netlib ({\tt http://\hskip0ex www.\hskip0ex netlib.\hskip0ex org}), a repository maintained
by
the AT \& T Bell Labs, the University of Tennessee, and the Oak Ridge National Laboratory.
Netlib provides access to thousands of pieces of software. Much of this
software is organized in the form of FORTRAN libraries.
For example, the QUADPACK library provides software routines for
the domain of numerical quadrature (the task of determining the
areas under curves, and the volumes bounded by surfaces).

\item {\bf Individual Recommender Systems:} These systems take a problem
description and identify a good algorithm, that satisfies user specified
performance constraints (such as error, time etc.).
For example, the GAUSS recommender system \cite{gauss} successfully selects algorithms
for numerical quadrature.
The issue of how to identify good algorithm recommendations
is a very complex and domain-specific one, and is
not covered here. We refer the reader to \cite{chirec,gauss} for more
details on this problem and a promising approach. It may be noted
that GAUSS uses collaborative filtering to correlate variations in
algorithm performance to specific characteristics of the problem input.
The patterns mined by GAUSS are relational rules that model the connection
between algorithms and the problems they are best suited to.

\item {\bf Cross-Indices of Software}: The GAMS (Guide to Available Mathematical Software) system
({\tt http://\hskip0ex gams.\hskip0ex nist.\hskip0ex gov}) provides a web based index for locating and identifying
algorithms for scientific computing. GAMS indexes nearly $10,000$ algorithms
for most areas of scientific software. While providing access to
four different Internet repositories, GAMS's main contribution to mathematical software,
however, lies in the tree structured taxonomy of mathematical and software problems used to
classify software modules. This taxonomy extends to seven levels and
provides a convenient interface to home in on appropriate modules.
Fig.~\ref{gams1} describes three screen shots during a GAMS session.
The top left part of Fig.~\ref{gams1} depicts the root GAMS node, where
the user is expected to make a selection about the type of problem
to be solved (arithmetic/linear algebra/quadrature etc.). The user
selects the `H' category and then proceeds to make further selections.
The class H2 corresponds to numerical quadrature, H2a corresponds to
one-dimensional numerical quadrature and so on. The H2a1
node is shown in the top right part of Fig.~\ref{gams1}. Continuing in this manner,
the `leaves' are arrived at (bottom part of Fig.~\ref{gams1}), where there still exist 
several choices of algorithms for a specific problem.
For some domains, it can be shown that there are nearly 1 million software
modules that are potentially interesting and significantly different
from one another! Recommenders for scientific software exist (as listed above) but
they work only in specific isolated subtrees of the GAMS hierarchy (like the GAUSS
system).

\end{itemize}

\noindent
We present an implementation of PIPE for personalizing recommendations about quadrature
software (GAMS category {\tt H2a}).
In the absence of PIPE, scientists typically use GAUSS to obtain a recommendation for a quadrature
algorithm, then manually navigate the GAMS taxonomy starting from the root, looking for the right
category containing an implementation for the recommended algorithm, and finally browse
the Netlib site to download the source code and documentation for the recommendation. It
is clear that any one of these resources does not provide enough information for
personalization. \\

\noindent
{\bf Experimental Setup:} Schema was first extracted from the Netlib site 
and personalized for the input (QUADPACK=1) (which provides the algorithms for
quadrature). The tree-building algorithm was written in Perl using
the navigation capabilities of the {\tt lynx} web browser. Perl hashes, which use
arrays indexed by page URL, helped identify commonalities in tree-building. The
mining algorithm did not yield any compression to the original Netlib schema for
QUADPACK, since it is a strict two-level hierarchy. A portion of the schema (simplified
for presentation) obtained
from Netlib is shown below:
\begin{center}
\begin {tabular}{|p{5in}|}\hline
\begin{verbatim}
    if (dqc25s.f)
         URL = "http://www.netlib.org/quadpack/dqc25s.f"
    ....
    if (readme)
         URL = "http://www.netlib.org/quadpack/readme"
\end{verbatim}
\\\hline
\end {tabular}
\end{center}
\vspace{.075in}

\noindent
Next, tree-building and data mining were conducted for
the GAMS website rooted at the {\tt H2a} node (one-dimensional numerical quadrature). Notice that while links 
in GAMS (and most web sites)
are not {\it typed}, we interpret
the text anchoring the `{\tt <a href>}'s in the web pages as the label when
following the associated link.
Furthermore, the labels for certain links (typically to software modules) are
very long that they cannot be listed intelligibly
on the originating page.
In such cases, the label is suffixed with
``...'' and continued on the page pointed to (see bottom part of Fig.~\ref{gams1}).
For the purposes of personalization, we cannot ignore the
continuation of the label as it may contain important keywords that describe
the module. The compressions arising from mining GAMS schema were of two main
flavors: (i) reductions due to factoring common nodes at the
pre-leaf level (typically module sets), and (ii) reductions
arising from links that violate the tree taxonomy. In overall, $80$ internal
nodes in the {\tt H2a} tree were reduced to $74$ nodes (after tree-building) and later,
to $69$ nodes (after data mining and collapsing multiple roles). Thus, a compression
of $14\%$ was observed for the {\tt H2a} GAMS subtree. The schema at this stage is
given by:
\begin{center}
\begin {tabular}{|p{5in}|}\hline
\begin{verbatim}
     if (Quadrature_Problem)
        if (One-Dimensional_Problem)
           if (Finite_Interval)
              if (Specific_Integrand)
                if (Automatic_Accuracy)
                  ....
\end{verbatim}
\\\hline
\end {tabular}
\end{center}
\vspace{.075in}
\noindent
where {\tt Quadrature\_Problem}, {\tt Automatic\_Accuracy} are the link labels
at the GAMS site. Finally, the recommendation rules from GAUSS \cite{gauss}
are already in programmatic form (they take a vector of
problem features and performance criteria
as input and make a recommendation for an algorithm):
\begin{center}
\begin {tabular}{|p{5in}|}\hline
\begin{verbatim}
   if (Int)
     if (Osc)
       if (finite)
          if (HighAcc)
             if (EndPtSing)
                algorithm = "Clenshaw-Curtis Quadrature"
             ....
      .....
\end{verbatim}
\\\hline
\end {tabular}
\end{center}
\vspace{.075in}

\noindent
The above three  schemas (programs) were merged taking into account the inconsistencies
in the labeling of the three web sources. For example, {\tt Int} in GAUSS is referred to
as {\tt Quadrature\_Problem} in GAMS, {\tt Finite} in GAUSS is cross-referenced 
as {\tt Finite\_Interval} in GAMS, and so on. 
The composite program was represented in the CLIPS programming language, which provides 
procedural, rule-based, and object-oriented paradigms for representation \cite{clips}.
The final program is structured as: 

\begin{center}
\begin {tabular}{|p{5.5in}|}\hline
\begin{verbatim}
main()
{
        /* assign feature values */
        /* code for matching variables that are cross-referenced */
        /* include program from GAUSS recommender system */
        /* include program from GAMS H2a website */
        /* include program for Netlib site */
}
\end{verbatim}
\\\hline
\end {tabular}
\end{center}
\vspace{.075in}

\begin{figure}
\centering
\begin{tabular}{cc}
& \mbox{\psfig{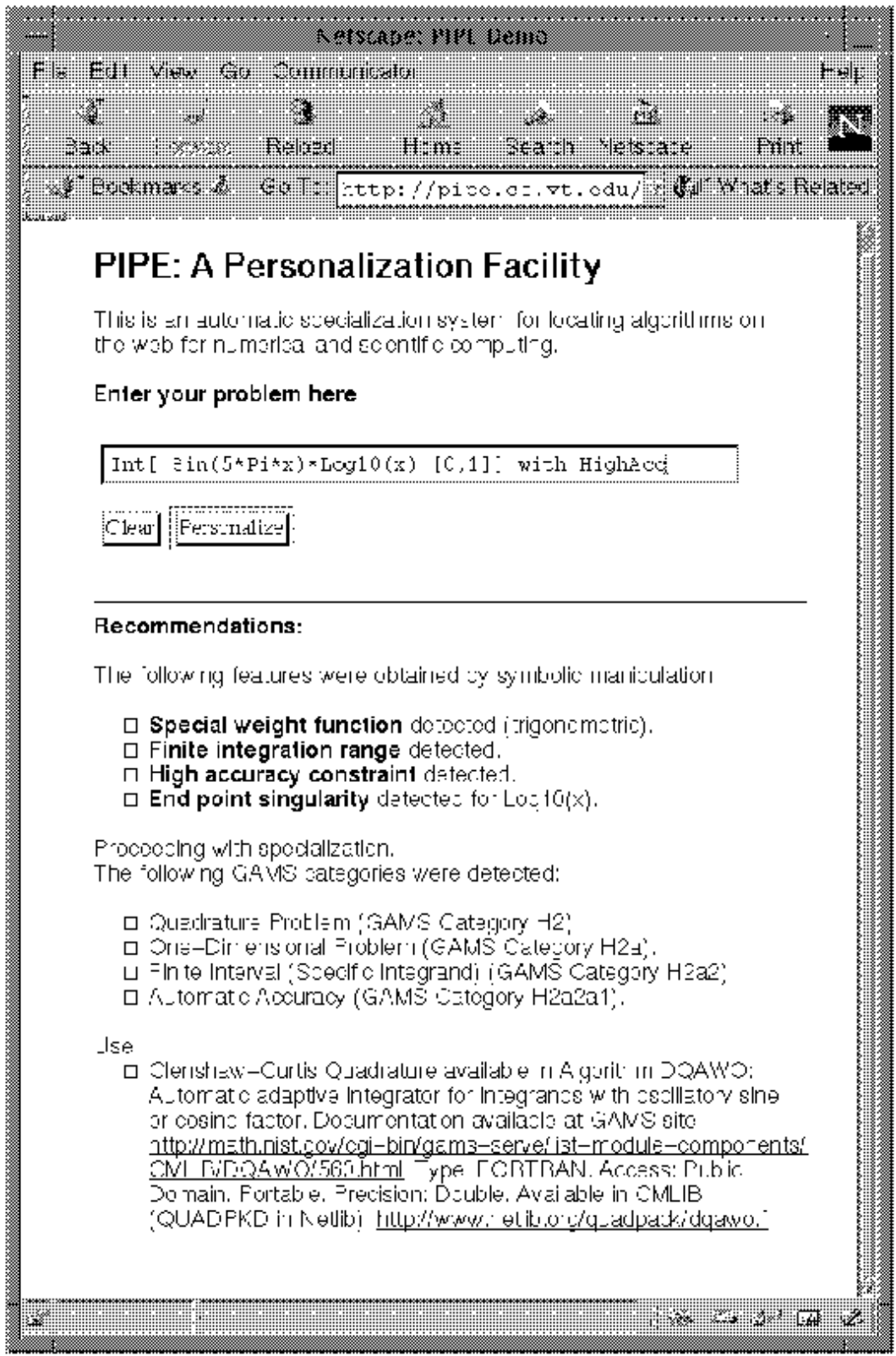}}\\
\end{tabular}
\caption{Sample session with the PIPE implementation for personalizing content for
a numerical quadrature problem.
Notice that the recommendation includes details of the algorithm
(and its implementation), the GAMS site
from where documentation is available, and the Netlib web repository from where the
source code can be downloaded.}
\label{person}
\end{figure}

\noindent
which models the control flow that gets partially evaluated.
The end-user interface for the
personalization system is shown in Fig.~\ref{person}. As shown,
the user provides the input
problem in self-describing mathematical terms. The implementation of
PIPE first parses the input to obtain as
many symbolic features as possible (it is to be noted that for this domain,
this involves some sophisticated mathematical reasoning). For instance,
in the example problem shown in Fig.~\ref{person}, simple parsing reveals
that the problem is a quadrature problem (from the presence of the {\tt Int}
operator) and that it is a one-dimensional problem (from the range restriction).
In addition, further mathematical reasoning reveals the presence of
an oscillatory integrand on a finite domain. For more details on how
this is achieved automatically, we refer the reader to \cite{gauss}.

Partially evaluating the CLIPS program above with this information by setting the appropriate
feature values to $1$ starts a cascading effect of program simplification, removing
nearly $95\%$ of the original information. The recommendation rules from GAUSS
get partially evaluated, in turn navigating the GAMS taxonomy rules, in turn
narrowing down on the Netlib URL for the selected algorithm. 
In this case, the evaluation is actually a {\it complete evaluation}, since the user
has provided enough information to zoom in on a final leaf. The resulting program is
then parsed to determine the 
individual program variables that are set at the end of this process. These are then
used to produce the output shown in Fig.~\ref{person} that includes (i) the
algorithm, (ii) the GAMS annotation, and (iii) the Netlib annotation indicating
the resource from which it can be downloaded.  The program segment producing the
output shown in Fig.~\ref{person} is given by:

\begin{center}
\begin {tabular}{|p{5in}|}\hline
\begin{verbatim}
    printout Algorithm "available in" GAMS_annotation 
    "Available in CMLIB (QUADPKD in Netlib)" URL;
\end{verbatim}
\\\hline
\end {tabular}
\end{center}
\vspace{.075in}

\noindent
Notice the use of the {\tt GAMS\_annotation} variable accompanying the node that was evaluated
which provides information about the documentation for the algorithm.
This implementation of PIPE should not be confused with a service run by
Wolfram, Inc. ({\tt www.integrals.com}) that evaluates quadrature problems symbolically.

\subsection{Personalizing the BEV}
The BEV ({\tt http://www.bev.net}) provides a
community resource for the New River Valley in Southwestern Virginia, USA,
where nearly $70\%$ of the population use the Internet actively
\cite{carroll-bev}. In its seventh year, BEV offers a wide array of services
--- information pertaining to arts, religion, sports, education, tourism,
travel, museums, health etc. An implementation
of PIPE was designed for this facility where the goal is to direct tourists
to appropriate resources in the town of Blacksburg. A first plan was
to use two resources --- the BEV web site (and various other pages that it links to)
and the Blacksburg Community Directory (an offshoot of the BEV site). We experienced
early problems with this approach due to ambiguities in the descriptions of the
BEV entities. Assume that a user is querying
for `art galleries.' Blacksburg boasts of nearly $25$ galleries; only $9$ of which
describe themselves as `galleries.' Others register their organizations
as `showrooms,' `centers,' or `museums' with the BEV site. \\

\begin{figure}
\centering
\begin{tabular}{cc}
& \mbox{\psfig{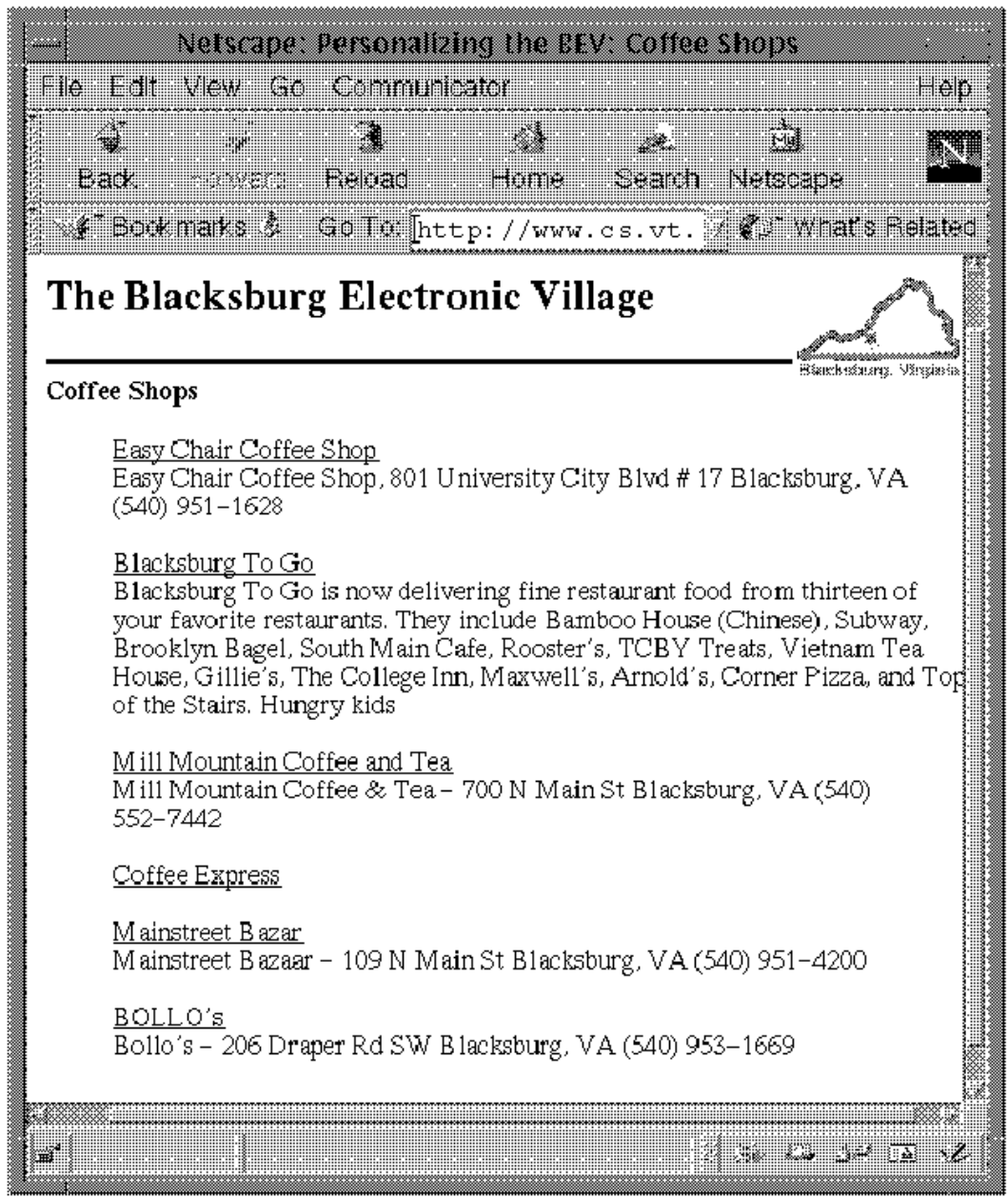}}\\
\end{tabular}
\caption{Personalizing the BEV for the query `Coffee Shops.' The addresses of
the various results are stored alongside the nodes as annotations and are reproduced
when nodes are selected in the final answer, as in the query above.}
\label{coffee}
\end{figure}

\noindent
{\bf Experimental Setup:} To overcome these problems, we introduced a third
personalization source --- TOPIC, a computational distillation of basic keywords and
topics used in the BEV site. The basic idea is to use orthogonal decompositions (such
as Singular Value Decompositions of the term-document matrix, or Lanczos decompositions)
to geometrically model semantic relationships. These approximations identify
hidden structures in word usage, thus enabling searches that go beyond simple keyword
matching. The exact computational algorithm is beyond
the scope of this article, but we refer the reader to papers such as \cite{textmining, terveen}
for this approach. The goal of TOPIC is to produce rules similar to the `Microsoft/MSFT'
matching that can be used to model recurrent low-dimensional subspaces in the BEV site(s).
The mining algorithm was applied to the BEV site but it did not 
yield as much a compression to
the original data as in the previous example. One reason for this could be the
lack of global `controls' in the construction of web pages by the BEV users and administrators.
However, partial evaluation did yield very effective results as shown in the
sample query of Fig.~\ref{coffee}. In this case, the evaluation is truly partial, since it
reproduces a collection of subtrees pertaining to `coffee.'
Notice that the second result depicted in
Fig.~\ref{coffee} is a `false positive,' since TOPIC associates the word `coffee' with
`cafe' whereas none of the resources identified in `Blacksburg to Go' are related to
coffee shops. A more detailed evaluation of both these case studies follows next.

\section{Evaluation of PIPE}
The above described implementations of PIPE involve substantially different
methodologies for evaluation. Consider the first case study, a domain of
importance and immediate relevance to the computational scientist.
Scientists and engineers would be experts at building
models in their particular domain, but novices at
understanding the intricacies of these mathematical models
and the software systems required to solve them. In other words, their
ratings/feedback {\it cannot} be used to evaluate PIPE's recommendations.
In addition, there does not exist any comparable comprehensive facility as described
in this article. This implementation of PIPE is hence a novel application of
personalization technology. To characterize the results, we used
a benchmark set of problems described in \cite{gauss} and applied (ran the algorithms,
effectively) the recommendations to see if they indeed satisfied the user's
constraints. In addition, we ensured that the web links from GAMS and Netlib were properly
associated with all the recommendations. 

First, all selections made by this implementation of PIPE were
`valid' (a selection is considered `invalid' if
the algorithm is inappropriate for the given problem, or if a wrong page from GAMS/Netlib
was indexed). In addition, we recorded the accuracy of the final selections
(a selection is accurate if the selected algorithm does
result in solutions satisfying the requested criteria). The best algorithm 
was selected 
for $87\%$ of the cases, and the
second best algorithm for $7\%$ of the cases. An
acceptable choice was made for $3\%$ of
the cases (which was not first or second best) and a wrong selection was made
for only
$3\%$ of the cases.
It is to be
mentioned that the `mistakes' arise from the uncertainties in
the GAUSS collaborative filtering system, and not as a result
of PIPE's methodology. Thus, it is seen that a
suitable recommendation is made most of the time.

The evaluation of the second implementation of PIPE is more tricky
and serves to illustrate the true potential of our methodology.
We adopted the following approach. $10$ people
were randomly selected and requested to identify $10$ queries (each)
that might be pertinent to a Blacksburg visitor (`hiking,' `mountains,' `trails' etc.).
We selected the $10$ most frequently cited queries
as test cases for PIPE's implementation. Stopword elimination (discarding terms
like `of' and `the' in queries) and stemming 
(`hikers' was mapped to `hiking,' for instance)
were first applied
to standardize these queries. 
These queries were then provided to $25$ Blacksburg
residents who were asked to enumerate the answers (from their point of
view) {\it before} personalization was conducted for these queries.
The results obtained were then provided as feedback and they were asked
if they would like to change their original answers or if they thought the results 
were deficient in any respect. Each user voted on a scale of 1--5 the mismatch
between the results and any `expected' answers (where a $1$ indicates that
he/she is completely satisfied with the results), for each of the queries.
We now summarize our results. 

First, all votes were in the range 1--2, with the
exception of $32$ votes with the value $3$ (more on this later). For each of the queries,
we then conducted a distribution-free test (the Kruskal-Wallis test \cite{statbook}) where
the hypothesis tested was that the results were unanimous versus the alternative that
they are not all equal. All $10$ hypotheses were accepted at the $95\%$ level, indicating
conclusively that the results were very close to the expected answers. The $32$ votes
with the value $3$ were spread over seven people who were less effusive with their
ratings than others. This is a standard problem with rating-based collaborative filtering;
one way to overcome this is to replace absolute ranks by `relative ranks' so that they
could be captured by certain two-way statistical tests (such as the Friedman, Kendall,
and Babington-Smith test \cite{statbook}). Another approach to overcome effusivity of
ratings (or the lack thereof) is presented in \cite{ira}. In addition, one of the
queries had consistently lower ratings by nearly all $25$ participants. This
was `Trails' which failed to reproduce two of the most popular trails in the vicinity
of Blacksburg. Not surprisingly, there were no web pages containing information about these
trails in the considered collection. The results also fared well when compared
with the traditional web search facilities available in the BEV site. For example,
the standard BEV search engine produced {\it no} results for
the query `coffee shops' (or coffee).

\section{Discussion and Comparisons with Other Approaches}
The effectiveness of the methodology presented above relies on
several factors, which are outlined below.

\begin{itemize}

\item The PIPE methodology allows programmatic composition to design full-fledged
systems and hence, comparisons with individual recommender systems or other
specialization facilities for particular domains are not strictly valid. With this in mind, 
one of its main advantages arises from integrating the design of 
personalization systems
with the task model(s) underlying the assumed interaction scenario. It is this property
that allows the designer to view the personalization system as a composition of individual
subsystems, using a programming metaphor. PIPE is hence restricted to those domains that
are most amenable to such decomposition and analysis techniques. More amorphous 
domains such as personalizing social networks in an organizational setting
might not fit this framework.

\item The above implementations assume that (i) the link labels
represent choices made by a navigator, and (ii) it is possible to ascertain the
values for such labels (program variables) from user input. In both cases
this is conveniently achieved, since the GAMS/GAUSS and BEV sites serve as 
ontologies that help guide the personalization process. 
While the notion of partial evaluation works for any site even in the 
absence of ontologies (as shown in {\it Example 1}), personalization will
only be as effective as the ease with which the link labels could be determined 
or supplied by the user. For example, if a medical informatics site is organized
according to scientific names of diseases and ailments, personalizing for `headaches'
will require a parallel ontology or cross-index that maps everyday words into
scientific nomenclature. In addition, the BEV case study shows that for certain domains
it is acceptable (or even desirable) to be less strict in variable
assignments, thus yielding more
false positives. For other domains, personalization might pose more stringent demands.

\item Personalizing textual content within web pages is not currently-addressed in the PIPE methodology
which relies on the accuracy of the links to point to appropriate information. Combining
textual mining techniques such as those presented in \cite{textmining} with the partial
evaluation concept is an interesting research issue for future investigation. 
Other approaches to content-based personalization arise from the
database management \cite{db-www} and information filtering \cite{lsi} communities. 
While languages like WebSQL, WebOQL, and Florid \cite{db-www} provide simple `web database' lookups, they
are not directly suited for personalization purposes since they accept queries in only a limited
form, and are more attuned to structure-querying across known levels.  However, such systems can
be used for program creation and associating augmented data structures with the program that
is partially evaluated.
In addition,
algorithms have been recently proposed that extract structure at the level of a single page.
These techniques infer Document Type Definitions (DTDs) and page schemas
from example web pages \cite{xtract}.
The integration of link-based analyses and such content-based schemes in a programmatic context 
also deserves exploration. 

\item In typical web sites, the links  are either mutually exclusive (e.g. in {\it Example 1}
and the GAMS case study), or are inclusive (e.g. the BEV case study). These are
currently modeled by the presence or lack of {\tt else if}s in our programs (respectively). 
This has the advantage of supporting both disjunctions and conjunctions in the personalization
queries.
Automating this aspect in a web crawler requires more study, e.g. via meta-data or
via explicit user direction. 

\item Partial evaluation, in general, is a costly operation because of the need to unroll loops
and complex control structures in programs. However, such features are almost always absent in
the kinds of studies considered here. Even links that point back to higher levels of the hierarchy do not
cause code blowups since they are factored by the mining process. As a result, the cost
of partial evaluation is not a severe bottleneck. 
The most expansive implementation of
PIPE is the first case study which involved sites that have tens of thousands of web pages.
Further analyses are needed, however, to characterize
the scale up with respect to an ever greater number of web pages. 

\item It is instructive to characterize how much of the encouraging results are obtained by
partial evaluation vis-a-vis carefully creating a tightly integrated implementation. In the
absence of handcrafting, partial evaluation will still lead to personalization, but 
such a scenario is highly unrealistic. Partially evaluating the program mined
from a beverages web site with respect to `Coke'
might not yield any results if the link label says `Coca-Cola.' Approaches to
alleviate this problem include the design of public ontologies and meta-data standards
for commercial domains of expertise. Until such a time when such ontologies become
prevalent on the web, this problem will not permit any general solutions.
In the absence of partial evaluation,
though, a more complicated strategy has to be in place to ensure that the personalization system
handles all possible types of queries, spanning all combinations of levels of the link labels.

\item Domain-specific techniques for making individual recommendations or choices
are not part of the PIPE framework {\it per se}
but will nevertheless form a critical aspect of
any successful personalization system. Various techniques have been proposed for
making recommendations; we refer the reader to \cite{ira,clever,lsi,grouplens,phoaks}.
The current methodology thus allows the designer to choose the rating
mechanism (value-neutrality). In the case of the first case study,
this is the GAUSS recommender system. A different system, trained
on qualitatively different examples, might be appropriate in another situation.

\item Little attention has been devoted toward automating the determination of
appropriate `starting points' for personalization. In the general case,
this should involve
a systematic way of finding authoritative sources. A good reference is \cite{clever}
which uses linear-algebraic matrix transformations to determine the most `cited' resources
(via hyperlinks) for a given topic.

\item The integrated methodology of PIPE is similar
in spirit to various other systems, most notably Levy's WebStrudel (a web site
management system) \cite{webstrudel}. Such systems 
typically take
a specification of a web site as input (graphs, rules etc.)
and define the structure of the site in terms of the underlying data model.
Currently, our implementations customize HTML content using the 
text manipulation
capabilities provided in Perl. Programmatic reconstruction of web pages
is a possible future extension of this work and systems like WebStrudel
can also eliminate
the need for restructuring when more web sources or additional rating
mechanisms are introduced.

\item In conclusion, the remark of Rus and Subramanian in \cite{rus} is especially pertinent:

\begin{descit}{}
``Whether users of the information superhighway prefer to
build their own `hot rods' [by methodologies like PIPE],
or take `public transportation' [web search engines]
that serves all uniformly is an empirical question and
will be judged by history.''
\end{descit}

\end{itemize}

\noindent
{\it Acknowledgements:} The author acknowledges the input of Sammy Perugini, Akash Rai,
Mary Beth Rosson and the nearly $40$ volunteers who helped in the evaluation of the second study. 
Feedback from several anonymous referees helped clarify the presentation and improve
the article.

\bibliographystyle{plain}
\bibliography{final}

\end{document}